\documentclass[12pt,a4paper]{article}
\usepackage[utf8]{inputenc}
\usepackage[T1]{fontenc}
\usepackage{amsmath}
\usepackage{amsfonts}
\usepackage{amssymb}
\usepackage{authblk}
\usepackage{graphicx}
\date{}
\title{Exploring Higher Dimensional Solutions for Conformal Cyclic Cosmology with DeSitter Space via Eternal Inflation}

\author[1,2]{Natarajan Shriethar \thanks{natarajangravity@gmail.com \\natarajan@spacequanta.com}}
\affil[1]{SpaceQuanta Lab, Tamilnadu, India}
\affil[2]{Qubitor lab, Singapore}

\begin{document}
\maketitle
\begin{abstract}
According to conformal cyclic cosmology, a two-surface (Penrose 2-surface) must exist between successive conformal aeons. To understand the quantum gravitational effects of the Penrose 2-surface, a few higher dimensional solutions are required.  The higher dimensional metric for the Penrose 2-surface is obtained in the current work.  The evolution of the conformal factor is discussed in this work. This article discusses the importance of higher dimensional solutions in understanding the final stages of the universe, particularly in the context of conformal cyclic cosmology. The article explores the behaviour of de Sitter space via eternal inflation, with the aim of providing a more complete understanding of the evolution of the universe and the role that higher dimensions may play in its final stages. The article concludes that without the availability of higher dimensions, the conformal birth of a new aeon will not be possible, and that the use of only three spatial dimensions in the final 2-surface of an aeon will fall short in accounting for the influence of quantum gravity. The article also presents a 5-dimensional metric that describes the behaviour of 5-dimensional spacetime, along with the corresponding 5-dimensional Friedmann-Robertson-Walker equations of the conformal surface in the late-time evolution of the universe. The results of this work have important implications for our understanding of the nature of space and time and how they behave in extreme conditions, such as the final stages of the universe.
\end{abstract}

\textbf{Keywords:}
    Higher Dimensional Conformal Solutions,Quantum Gravity, Eternal Inflation,Conformal Cyclic Cosmology, Big Rip
\section{Introduction}

The big bang model is a leading model for understanding the evolution of the universe. But the big bang model lacks the answers to the questions like, what is beyond the big bang singularity? what will happen beyond the future big crunch? To resolve these questions the conformal cyclic model (CCC) is proposed \cite{penrose2012basic,penrose2010cycles}.

Conformal Cyclic Cosmology (CCC) is a cosmological theory which describes the universe as being cyclic, where each cycle starts with a Big Bang and ends with a Big Crunch, and it is conformally related to the previous cycle. The theory proposes that the final stages of each cycle are described by higher dimensional solutions to the Einstein field equations, which allow for the transfer of information from one cycle to another through conformal rescaling. The CCC theory differs from traditional cosmological models in that it avoids the initial singularity and smooths out the transition from one cycle to the next. It is based on the idea of conformal geometry, which deals with the behavior of shapes and sizes under changes in scale, and the conformal invariance of the laws of physics. Despite being an intriguing idea, CCC remains a controversial and highly speculative theory that requires further observational evidence to be considered a credible explanation of the universe's evolution.

The CCC model explains the evolution of the universe in terms of successive conformal cycles (namely aeons). The model connects the evolution of two aeons via a 2 surface, (to be named as Penrose 2 surface in the present work). The hypersurface connects the two successive singularities, by introducing topological contraction and expansion. The conformal cyclic cosmological model \cite{penrose2014gravitization} explains the evolution of the universe in a modified fashion rather than orthodox big bang evolution. 

Penrose 2-surface is the connecting surface between two aeons  \cite{araujo2015spacetime}.
Recently the model gains its first success with experimental results. With the available cosmic microwave background data, the consequences from the previous aeon in the CMB spectrum are reported by Penrose and his team \cite{an2020apparent}. While the transition from one aeon to another aeon, the quantum gravitational effects are necessary for the beginning of the future aeon.

The Penrose 2 surface, also known as the conformal 2-surface, is a mathematical concept that is used to study the structure of the universe. It is a two-dimensional surface that connects two successive cycles of the universe, providing a way to analyze the geometric structure of space-time. The Penrose 2 surface is an important concept in the field of cosmology and theoretical physics, as it provides insight into the nature of the universe and its evolution.

The Penrose 2 surface plays a crucial role in understanding the structure of the universe. It is considered as a tool to analyze the geometric structure of space-time, which is the fundamental structure of the universe. The study of the Penrose 2 surface allows scientists to examine the relationships between the physical and mathematical properties of the universe, and to explore the nature of gravity and the behavior of matter and energy.
The Penrose 2 surface is conformal, meaning that it preserves angles and shapes, but not necessarily sizes. This property makes it an important tool in the study of the universe, as it provides a way to analyze the geometric structure of space-time in a way that is independent of the scales involved.

In addition, the Penrose 2 surface is also significant in the study of black holes and the formation of singularities. Singularities are regions of space-time where the laws of physics break down, and the Penrose 2 surface provides a way to understand these regions and the effects they have on the surrounding space-time.
The Kaluza-Klein theory is a proposed unification of gravity and electromagnetism in a higher-dimensional space. It postulates that the universe has more than four dimensions, and that the extra dimensions are curled up and hidden from us \cite{freund1982kaluza,branding2019stable}. This theory can be used to explain why some particles have properties that are difficult to understand in four-dimensional spacetime.

Meanwhile, the quantum gravitational effects are eloquently studied by loop quantum gravitational (LQG) formalism \cite{thiemann2003lectures}. Formalism sheds light on the fundamental structure of spacetime. The model predicts the discrete nature of spacetime in quantum geometry. The LQG theory resolves the geometry of curvature singularities that emerges in various scenarios in the cosmos. The black hole singularity \cite{modesto2004disappearance}, big bang singularity \cite{ashtekar2006quantum} and even higher-order singularities \cite{cailleteau2008singularities, ashtekar2021short, natarajan2019resolution, sami2006avoidance} are also resolved by LQG formalism. The loop quantum cosmological principle \cite{agullo2014loop, ashtekar2009loop} is inspired by the loop quantum gravity formalism \cite{perez2017black}.

Phantom energy is a mysterious component of the universe. The phantom energy is a hyper form of dark energy that has the equation of state $w<-1$. In the late time scenario, the universe will be dominated by phantom energy. The phantom energy causes accelerated expansion. The domination of phantom energy rip off the whole universe which will cause a future big rip. The strength of the phantom energy can separate the galactic clusters, galaxies, stars and even subatomic particles \cite{caldwell2002phantom,caldwell2003phantom}. The presence of such phantom energy is a vital role in the phantom-dominated conformal cyclic cosmology to have higher dimensional solutions.

By fusing the conformal cyclic cosmology with loop quantum cosmology, in the big rip background and necessity of higher dimensions are proposed elsewhere \cite{ours2}. From such a viewpoint the dynamics of the aeonic connecting 2 surfaces can be studied. Also, the higher dimensional necessity required on Penrose 2 surface is to implement the quantum gravitational effects. Such implications are also studied in the current work.  The modified Raychaudhuri equations from the braneworld are obtained. Also, the cosmological constant from loop quantum solutions and the AdS/CFT perspective are bridged. The brane tension is exclusively calculated using loop quantum cosmology and AdS/CFT perspective. The topological expansion and contraction solutions of the conformal cyclic evolution are mapped with the braneworld-like scenario.
The obtained solutions are verified with SageManifolds computer package \cite{birkandan2019symbolic, gourgoulhon2015tensor}.  The conformal factor is discussed with its differential form for the Penrose 2-surface. Similarly, the stress-energy tensor of the Penrose 2-surface is also calculated and verified.

The section \ref{hdcp} appears to be discussing a mathematical derivation involving higher-dimensional Penrose 2-surfaces, conformal metrics, and stress-energy tensors.

Section \ref{rhds}  discusses the need for higher-dimensional solutions in the conformal cyclic cosmology theory, which predicts the end and beginning of the universe merging at a Penrose 2-surface. This surface requires both quantum gravitational and classical mechanical properties, but the three spatial dimensions in the final 2-surface are not enough to explain the effects of quantum gravity on the initial singularity of the next aeon. Therefore, higher dimensional solutions and big bounce solutions are necessary to fuse the expanded and shrunk dual 2-surfaces. The presence of phantom energy is considered to understand such evolution, as it rips off every constituent of the universe and leads to a big rip. The article also mentions the need for DeSitter solutions in a scenario where the late-time universe may face a big rip due to dark energy called phantom energy. The article also proposes using the braneworld scenario to connect the loop quantum evolution with string theory solutions and uses a 5-dimensional brane metric to map the 3+1D universe into the braneworld scenario.

Eternal inflation is a theory in modern cosmology that suggests that the universe is expanding at an accelerating rate and will continue to do so forever \cite{guth2007eternal,borde1994eternal,kinney2019eternal}. This idea is a natural consequence of the inflationary theory proposed in the 1980s to explain several puzzles in the early universe, such as the flatness and horizon problems. Inflation suggests that the universe underwent an exponential expansion phase in its early moments, which flattened it out and smoothed out any irregularities. However, eternal inflation takes the concept further by proposing that the inflationary phase never truly ended and that the universe is continually expanding, with new pockets of inflation occurring indefinitely. This theory has profound implications for our understanding of the origin and fate of the universe, and it continues to be an active area of research and debate in cosmology.

Hence the section \ref{etinf} explains the possible consequence between the conformal parameter and the eternal inflation. This section discusses the details of the Kaluza-Klein metric and the conformal metric, and their applications in the context of eternal inflation and conformal cyclic cosmology.

The Penrose 2-surface expresses the existence of duality between the quantum gravitational solution and general relativity solutions. By understanding its mathematical analogy, the conformal extension of the universe and the conservations of the information through the successive aeons can be understood. 

Some solutions like the Ekpyrotic model \cite{steinhardt2002cosmic} offer curvature-free evolution of the universe. From that viewpoint, it is also predicted that higher-dimensional evolution may be possible in another dimension too.

\section{Requirement of Higher dimensional solutions}\label{rhds}
Higher dimensional solutions are important on conformal Penrose diagrams because they provide a way to understand the behavior of gravity in the final stages of the universe, when it is expected to be dominated by the effects of quantum gravity. In this regime, traditional 4-dimensional solutions to Einstein's equations may not be sufficient to describe the physics accurately, and higher dimensional solutions can provide a more complete picture. These higher dimensional solutions can also help to resolve some of the challenges faced by current models of the universe, such as the cosmological constant problem and the black hole information paradox. By considering these higher dimensional solutions, scientists can gain a deeper understanding of the nature of space and time, and how they behave in the extreme conditions of the final stages of the universe.

The CCC theory suggests that the final stage of each cycle is described by a de Sitter space, which is a space with positive cosmological constant and exponential expansion.

In this article, you aim to solve the conformal 2 surface with de Sitter space via eternal inflation. Eternal inflation is a process in which certain regions of the universe undergo exponential expansion due to a positive cosmological constant. This expansion is thought to be driven by a scalar field, known as the inflaton, that has a large positive vacuum energy. The idea behind eternal inflation is that once the inflaton field begins to roll down its potential, it produces regions of space where the energy density is high enough to drive exponential expansion.

The goal of solving the conformal 2 surface with de Sitter space via eternal inflation is to provide a more complete understanding of the evolution of the universe and the role that higher dimensions may play in the final stages of its evolution. By exploring the mathematical and physical properties of this system, you hope to shed light on the mechanisms that drive eternal inflation and the behavior of the universe in the final stages of each cycle. Ultimately, this work could lead to a deeper understanding of the universe and its ultimate fate.

The conformal cyclic cosmology predicts the end evolution of the universe and the beginning of the universe merges at the Penrose 2-surface. Such Penrose 2-surface would behave like a geometrically coupled expanded surface. Similarly, geometrical manipulation of such surfaces requires both quantum gravitational and classical mechanical properties. As it is an expanded classical 2-surface, the spacetime behaves as Minkowski flat metric. But the beginning of the universe requires high curvature and quantum gravitational properties. 

However, the utilization of only three spatial dimensions in the final 2-surface of an aeon will fall short in accounting for the influence of quantum gravity that addresses the starting singularity of the next aeon. Hence to fuse the expanded and shrank dual 2-surface, the quantum gravitational solutions with classical space-time are meant to be necessary. Without the availability of higher dimensions, the conformal birth of a new aeon will not be possible as the expanded space misses the quantum gravitational effects. The higher dimensional solutions and big bounce kind of repulsive solutions, there require an exciting form of energy in the Penrose 2-surface.

To understand such evolution the presence of phantom energy is considered. The phantom energy rips off every constituent of the universe and lets it end with a big rip. Here for the conformal aeons, the phantom energy let the universe extend even beyond the final singularity and induces the quantum gravitational solutions in the Penrose 2-surface.

The classical Minkowski space solutions contain negligible cosmological constant. But the later universe might consist of a non-zero cosmological constant. The late time expanded universe may even face the big rip by the consequence from the form of dark energy called phantom dominated scenario \cite{deu2}. The phantom-dominated conformal evolution of the universe is discussed in \cite{ours1}. These solutions concluded that there would be the higher dimensional evolution in order to have conformal continual evolution of the universe. Similarly for such a scenario the DeSitter solutions are also required.

The conformal factor and extension of the universe beyond any future singularities are proposed in \cite{shriethar2020conformal,natarajan2022cyclic}.

Hence for mapping of the 3+1 D universe into the extended conformal scenario, a 4+1 D metric is considered. 

The 5-Dimensional  metric is written as \cite{ours2}.
\begin{equation}\label{met1}
\begin{split}
ds^2 = -dt^2 + \left( -\frac{a\left(t\right)^{2}}{k r^{2} - 1} \right) dr^2 + r^{2} a\left(t\right)^{2} d \theta^2 + r^{2} a\left(t\right)^{2} \sin\left({\theta}\right)^{2} d \phi^2 \\ + \left( -k r^{2} + 1 \right) d \chi^2
\end{split}
\end{equation}

The metric in equation \ref{met1} defines the geometry of a spacetime, in this case, it describes the behavior of 5-dimensional spacetime.

The 5-dimensional FRW equations  are obtained as

\begin{equation}\label{frw1}
H^2 = \frac{8 \pi G \rho}{3} - \frac{2 k}{a^2} + \frac{\Lambda}{3}
\end{equation}

The above equation (\ref{frw1}) represents the 5-dimensional Friedmann-Robertson-Walker (FRW) equations of the universe in the late-time evolution of the conformal universe. The higher dimensional solutions are plausible during the late-time evolution of the conformal universe to preserve the quantum gravitational effects. The DeSitter space with quantum gravitational effects will act on the Penrose 2-surface, which will help to determine the validity of the laws of physics on the conformal translation.

Before analysing the conformal evolution further, let us glance the quantum cosmological analysis of the final stages of the universe. 

To understand the quantum cosmological evolution in the late time universe scenario, we acn see the modified FRW equation from the loop quantum evolution. \cite{bojowald2008loop}

\begin{equation}\label{frw1.1}
H^2 = \frac{8 \pi G \rho}{3}\left(\frac{1- \rho}{\rho_c} \right) - \frac{2 k}{a^2} + \frac{\Lambda}{3}
\end{equation}
Here $\rho_c$ is the critical density \cite{ashtekar2003mathematical}. It has the values in the order of $\rho_c \sim 0.41 \rho_{pl}$.  The bounce will occur at $\rho \to \rho_c$. That bounce will be in higher dimensions as the corresponding metric is said to have higher dimensions.  Similarly  the modified Raychoudri equation in terms of loop quantum cosmology is obtained as,
\begin{equation}\label{frw2}
\frac{\ddot{a}}{a} = -\left(\frac{4 \pi G}{3} (4P+\rho)-\frac{\Lambda}{3}\right)
\end{equation}

The equation \eqref{frw2} represents the acceleration of the scale factor $a$ of the universe in terms of its energy density $\rho$, pressure $P$, and cosmological constant $\Lambda$.

As the equation of state is written as $w=\frac{p}{\rho}$ and for dark energy $w=-1$. Hence equation \ref{frw2} is described as 

\begin{equation}\label{frw2.a1}
\frac{\ddot{a}}{a} = 4 \pi G \rho + \frac{\Lambda}{3}
\end{equation}

The positivity of the equation \ref{frw2.a1} confirms the exponential expansion of the universe. Hence in the late time, with more postive cosmological constant, the universe will continue its expansion without being perturbed by inbound gravitational pull.

As suggested the phantom dominated final stages of an aeon will have the equation of the state as  $w <-1$. For a simple case the equation of state is taken as $w=-1.5$. Hence the equation \ref{frw2} is written as 
\begin{equation}\label{drk2}
\frac{\ddot{a}}{a} = \frac{\Lambda+ 20 \pi G \rho}{3}
\end{equation}

The phantom energy with $w <-1$ let the expansion of the universe in an accelerative manner. Equation \ref{drk2} indicates the phantom energy dominated accelerated expansion of the universe. As said earlier the phantom energy is a cause for the hyper interaction which leads to a big rip. The equation leads the evolution towards the big rip singularity.

The modified Raychoudri equation can be obtained for the equation \ref{frw2}, to get the acceleration in quantum cosmological regime. Hence
\begin{equation}\label{frw2.b1}
\frac{\ddot{a}}{a} = -\left(\frac{4 \pi G}{3} \left(4P \left(1- 2 \frac{\rho}{\rho_c}\right)+\rho \left(1- 4 \frac{\rho}{\rho_c}\right)\right)-\frac{\Lambda}{3}\right)
\end{equation}

The LQC-powered Raychouduri equation corresponds to higher dimensional evolution by the turnaround points at $\rho \to \rho_c$. As said earlier, the phantom-dominated big rip can not be avoided by equation \ref{drk2}. But the modified equation in \ref{frw2.b1}, avoids the future big rip by attaining the critical density. The LQC plays an important role in these kinds of bounces and it is also found that the beyond bounce expansion will be in higher dimensions. Hence the non-stoppable conformal evolution even beyond the big rip is also available as a part in the confrmal evolution of the universe.

Further from the loop quantum gravitational solutions, the cosmological constant can be reconciled with quantum gravitational resolutions as suggested from \cite{mizoguchi1992three}
\begin{equation}\label{sc3}
\Lambda = \frac{\pi}{j_{max} \hbar G}
\end{equation}
By substituiting equation \ref{sc3} into \ref{frw2.b1} the modified Raychoudri equation is obtained with quantized cosmlogical constant as
\begin{equation}\label{frw2.b2}
\frac{\ddot{a}}{a} = -\left(\frac{4 \pi G}{3} \left(4P \left(1- 2 \frac{\rho}{\rho_c}\right)+\rho \left(1- 4 \frac{\rho}{\rho_c}\right)\right)-\frac{\pi}{3 j_{max} \hbar G} \right)
\end{equation}

As $j$ approaches $j_{max}$ the quantum evolution retains the classical regime. Hence the full theory of quantum cosmology is obtained. Hence equation \ref{frw2.b2} predicts the evolution in higher dimensional conformal cyclic perspective in the view of loop quantum cosmology which offers non-singular bounce.

As the quantum number $j$ approaches the maximum value $j_{max}$, the universe's evolution moves from a quantum to a classical regime. This equation predicts the evolution of the universe in higher dimensions, in the view of loop quantum cosmology, which offers a non-singular bounce solution. This equation is important as it provides insight into the behavior of the universe in the quantum cosmological regime.

\section{Higher Dimensional FRW solutions}

In the previous section, we discussed the 5-dimensional FRW solutions and their implications in understanding the evolution of the universe. In this section, we extend our analysis by introducing an additional dimension and examining the consequences of this extension on the standard FRW and Raychoudri equations. By incorporating the principles of loop quantum cosmology, we derive modified versions of these equations and explore their implications in avoiding future singularities and the big rip. The modified solutions provide a quantum cosmological perspective on the evolution of the universe in higher dimensions.

Considering the higher dimensional geometrical surface, the modified resolution can be obtained for the FRW equation with LQG implementation as,
\begin{equation}\label{frw4}
\begin{split}
\left(\frac{\dot{a}}{a}\right)^2= \frac{8 \, \pi G \rho\left(t\right)}{3}\left(1-\frac{\rho}{\rho_c}\right) + \frac{\Lambda}{3} 
- \frac{23 \, k}{12 \, a\left(t\right)^{2}} + \frac{1}{3 r^{2} a\left(t\right)^{2}} + \frac{1}{12 \, r^{2} a\left(t\right)^{2} \sin\left({\theta}\right)^{2}} \\ - \frac{1}{12 \, {\phi}^{2} r^{2} a\left(t\right)^{2} \sin\left({\theta}\right)^{2}} 
\end{split}
\end{equation}

The equation \ref{frw4} describes the modified Friedmann-Robertson-Walker (FRW) solution in a higher-dimensional universe. It states that the expansion rate of the universe (represented by $\frac{\dot{a}}{a}$) is determined by various factors such as the density of the universe at time $t$ (represented by $\rho(t)$), the cosmological constant ($\Lambda$), and geometric terms. The equation also includes the density threshold ($\rho_c$) at which the universe will bounce due to quantum gravitational effects. This bounce is expected to have a generic character in the higher-dimensional universe.

On larger scales, the requirement of higher dimensions requires the necessity for the quantum gravitational solutions. Hence such implications of quantum gravitational properties on the higher dimensions will led them to implement on Penrose 2 surface. As the loop quantum solutions resolve the singularity, the consequencial topological conformal 2 surface will also behave in such the way. In addition to these, it is also be known as the DeSitter resolution gives another freedom to have the presence of cosmological constant in the evolution of the universe.

The Raychoudri equation for the six-dimensional  manifold is represented as

\begin{equation}\label{frw5}
\frac{\ddot{a}}{a}= \frac{-20 \, \pi G p\left(t\right)}{3} - \frac{4 \, \pi G \rho\left(t\right)}{3} +\frac{\Lambda}{3} 
\end{equation}
\begin{equation}\label{frw5.1}
\frac{\ddot{a}}{a}= \frac{-4 \pi G}{3} \left( 5P+ \rho \right) +\frac{\Lambda}{3}
\end{equation}

In a six-dimensional manifold, Raychoudri equation is represented by equations \ref{frw5} and \ref{frw5.1}. Equation \ref{frw5} expresses the change in the second derivative of the scale factor ($\ddot{a}/a$) with respect to time. On the right side of the equation, it includes terms representing the effects of pressure ($p$) and density ($\rho$) on the expansion rate, as well as the cosmological constant ($\Lambda$). Equation \ref{frw5.1} presents a similar expression, but with the pressure and density terms combined as 5 times the pressure plus the density. The equation is important for understanding the role of the density and pressure of matter and energy in the universe in driving its expansion or contraction.
By setting the equation of state $w \sim -1.5 $ for the phantom energy, the equation can be modified  as 

\begin{equation}\label{frw5.2}
\frac{\ddot{a}}{a}=   +\frac{26 \pi G+\Lambda}{3}
\end{equation}

The equation of state $w \sim -1.5 $ represents the nature of the phantom energy, which is a form of dark energy with negative pressure. By incorporating this equation of state into the Raychoudhuri equation, the modified equation \ref{frw5.2} is obtained. This equation represents the rate of change of acceleration of the six-dimensional manifold with a term that includes both the cosmological constant $\Lambda$ and 26 times the gravitational constant $G$. The value of $w \sim -1.5$ has significant implications for the evolution of the universe, as it represents a form of energy that causes the universe to expand at an increasing rate.

The equation \ref{frw5.2}  tells the accelerated expansion of the universe over larger scales. The equation remains positive due to the presence of phantom energy. The universe eventually faces a big rip by having positivity in equation \ref{frw5.2}

The equation \ref{frw5} can be modified with the quantum version of the Rayuchoudri equation. The modified Raychoudri equation is obtained for this evolution as

\begin{equation}\label{frw5.a1}
\frac{\ddot{a}}{a}= \frac{-4 \pi G}{3} \left( 5P\left(1- 2 \frac{\rho}{\rho_c}\right)+ \rho \left(1-4 \frac{\rho}{\rho_c} \right) \right) +\frac{\Lambda}{3}
\end{equation}

The equation \ref{frw5.a1} represents a modified version of the Raychoudri equation that takes into account quantum effects. The equation is obtained by the inclusion of terms that depend on the density $\rho$ and its critical value $\rho_c$. These terms capture the behavior of the universe in a higher dimensional and quantum gravitational perspective. The equation \ref{frw5.a1} can be used to describe the evolution of the universe at the quantum level.

The previously predicted big rip is avoided in this evolution as the universe attains critical densities. Thus any higher dimensional evolution in the conformal universe will avoid the singularities and they evolve continuously by the loop of quantum cosmological resolutions.  Hence it is also been understood that the conformal quantum universe has the inevitable presence of a cosmological constant. The modified cosmological constant is introduced in the equation as

\begin{equation}\label{frw5.a2}
\frac{\ddot{a}}{a}= \frac{-4 \pi G}{3} \left( 5P\left(1- 2 \frac{\rho}{\rho_c}\right)+ \rho \left(1-4 \frac{\rho}{\rho_c} \right) \right) +\frac{\pi}{3 j_{max} \hbar G}
\end{equation}

The equation \ref{frw5.a2} provides a modified Rayachoudri equation with the quantized cosmological constant. As the phantom energy available in the late time evolution, the universe will continue its expansion and the Penrose 2 surface will have the topological evolution in the quantum gravity scales. 

\section{Evolution of the conformal parameter}\label{hdcp}

Among the conformal cyclic evolution, the brane cosmology implements the evolution beyond the end point of evolution.  Such conformal transformation is studied with conformal parameters. Regarding these transformation, $AdS_{D+1}$ is considered. 

As suggested earlier the Minkowski manifold in the Penrose 2-surface requires additional dimensions to cope with quantum gravitational solutions. Hence in Minkowski spacetime, the 5-dimensional manifold is considered.

The metric in this manifold  is defined in (from equation \ref{met1}).

The null coordinates are expressed as

\begin{eqnarray} u & = & -r + t \\ v & = & r + t \\ {\theta} & = & {\theta} \\ {\phi} & = & {\phi} \\ {\chi} & = & {\chi} 
\end{eqnarray}

Here, $u$ and $v$ are light-cone coordinates, which means that light rays move along lines that are at a 45-degree angle to these axes. The coordinate $t$ is time, and $r$ is a radial coordinate. The coordinates ${\theta}$, ${\phi}$, and ${\chi}$ are angular coordinates.

The use of null coordinates is particularly helpful for describing the behavior of light and gravity in the vicinity of strong singular region, where the curvature of spacetime is very strong. In these coordinates, the metric that describes spacetime takes on a simple form, which allows for easier calculations and insights into the physics of singularities.

The inverse of the null coordinate transformation expresses the original spacetime coordinates, namely $t$, $r$, $\theta$, $\phi$, and $\chi$, in terms of the null coordinates $u$ and $v$.

Its inverse is written as 
\begin{eqnarray} t & = & \frac{1}{2} \, u + \frac{1}{2} \, v \\ r & = & -\frac{1}{2} \, u + \frac{1}{2} \, v \\ {\theta} & = & {\theta} \\ {\phi} & = & {\phi} \\ {\chi} & = & {\chi} 
\end{eqnarray}

Hence the higher dimensional metric  is derived as,
\begin{equation}\label{key}
\begin{split}
ds^2 = -\frac{1}{2} du dv -\frac{1}{2} dv du + \left( \frac{1}{4} \, u^{2} - \frac{1}{2} \, u v + \frac{1}{4} \, v^{2} \right) d \theta^2 \\ + \left( \frac{1}{4} \, u^{2} \sin\left({\theta}\right)^{2} - \frac{1}{2} \, u v \sin\left({\theta}\right)^{2}  + \frac{1}{4} \, v^{2} \sin\left({\theta}\right)^{2} \right) d \phi^2 \\+ \left( -\frac{1}{4} \, k u^{2} + \frac{1}{2} \, k u v - \frac{1}{4} \, k v^{2} + 1 \right) d \chi^2
\end{split}
\end{equation}

The conformal metric is introduced for the bluk evolution \cite{shriethar2020conformal},

\begin{equation}\label{cf2}
\tilde g = \Xi^2 g
\end{equation}

In the context of general relativity, the conformal metric is a new metric defined in terms of the original metric. In the equation you provided, $\tilde g$ represents the conformal metric, and $g$ is the original metric.

The conformal metric is introduced to simplify calculations in the  evolution of a system. The scaling factor $\Xi$ is a function that depends on the properties of the system, and is used to scale the original metric $g$. The conformal metric $\tilde g$ is obtained by multiplying the original metric $g$ by the square of the scaling factor $\Xi$.

The introduction of the conformal metric can simplify the analysis of certain systems in general relativity, as it allows for the use of conformal transformations, which can reveal important properties of the system. The choice of the scaling factor $\Xi$ depends on the specific system being analyzed, and is often chosen to simplify the equations of motion.


The conformal factor can be written in a differential form.

\begin{equation}\label{cf5}
\mathrm{d}\Xi = \left( -\frac{\varsigma}{\vartheta} \right) \mathrm{d} t + \left( -\frac{\varpi}{\varphi} \right) \mathrm{d} r
\end{equation}

In the context of the equation \ref{cf2} the conformal factor $\Xi$ is a scalar function that rescales the metric $g$. However, in some cases, it can be useful to express the conformal factor in terms of differential forms, which are mathematical objects that generalize the concept of a function.

The equation \ref{cf5} expresses the conformal factor as a 1-form, which is a linear combination of the differentials of the independent variables $t$ and $r$, with coefficients $\left( -\frac{\varsigma}{\vartheta} \right)$ and $\left( -\frac{\varpi}{\varphi} \right)$, respectively. This means that the value of the conformal factor at any point in space-time is determined by the values of the coefficients and the differentials of $t$ and $r$ at that point.

The values of $\varsigma$ and $\vartheta$ are given by the expressions

Here $\varsigma = 4 \, {\left(t^{3} - {\left(r^{2} - 1\right)} t\right)} \sqrt{r^{2} + 2 \, r t + t^{2} + 1} \sqrt{r^{2} - 2 \, r t + t^{2} + 1}$ \\and $\vartheta = r^{8} + t^{8} - 4 \, {\left(r^{2} - 1\right)} t^{6} + 4 \, r^{6} + 2 \, {\left(3 \, r^{4} - 2 \, r^{2} + 3\right)} t^{4} + 6 \, r^{4} \\- 4 \, {\left(r^{6} + r^{4} - r^{2} - 1\right)} t^{2} + 4 \, r^{2} + 1$. \\ 
The values of $\varsigma$ and $\vartheta$ are used in this integration process to obtain the value of $\Xi$ at a given point.

$\varpi = 4 \, {\left(r^{3} - r t^{2} + r\right)} \sqrt{r^{2} + 2 \, r t + t^{2} + 1} \sqrt{r^{2} - 2 \, r t + t^{2} + 1}$ \\

$\varpi$ is a function of $r$ and $t$. It is defined in terms of two square roots and a polynomial. The first square root is $\sqrt{r^{2} + 2 , r t + t^{2} + 1}$, and the second square root is $\sqrt{r^{2} - 2 , r t + t^{2} + 1}$. The polynomial in the expression for $\varpi$ is $r^{3} - r t^{2} + r$.

Similarly in  $\varphi = r^{8} + t^{8} - 4 \, {\left(r^{2} - 1\right)} t^{6} + 4 \, r^{6} + 2 \, {\left(3 \, r^{4} - 2 \, r^{2} + 3\right)} t^{4} + 6 \, r^{4} \\- 4 \, {\left(r^{6} + r^{4} - r^{2} - 1\right)} t^{2} + 4 \, r^{2} + 1$ \\ $\varphi$ is also a function of $r$ and $t$, and it is defined in terms of several polynomial terms of $r$ and $t$.

Interestingly the conformal factor becomes, 
\begin{equation}\label{cf9}
\mathrm{d}\Xi = -2 \, \cos\left(\frac{1}{2} \, {\tau}\right) \sin\left(\frac{1}{2} \, {\tau}\right) \mathrm{d} {\tau} -2 \, \cos\left(\frac{1}{2} \, {\Omega}\right) \sin\left(\frac{1}{2} \, {\Omega}\right) \mathrm{d} {\Omega}
\end{equation}

The equation \ref{cf9} is showing an alternative representation of the conformal factor in terms of two new variables $\tau$ and $\Omega$, where the differential form $\mathrm{d}\Xi$ is given in terms of the differentials $\mathrm{d}\tau$ and $\mathrm{d}\Omega$.

The equation shows that the conformal factor can be expressed in a trigonometric form involving the sine and cosine functions. The $\tau$ and $\Omega$ terms in the equation are related to the original coordinates $t$ and $r$.

The presence of the sine and cosine functions indicates that the conformal factor may have some periodicity or oscillatory behavior with respect to the new variables $\tau$ and $\Omega$.

Hence the higher dimensional  conformal metric is redefined as

\begin{equation}\label{cf10}
\begin{split}
\tilde{g} = \left( -\frac{2}{{\left(u^{2} + 1\right)} v^{2} + u^{2} + 1} \right) du dv + \left( -\frac{2}{{\left(u^{2} + 1\right)} v^{2} + u^{2} + 1} \right) dv du \\+ \left( \frac{u^{2} - 2 \, u v + v^{2}}{{\left(u^{2} + 1\right)} v^{2} + u^{2} + 1} \right) d \theta^2 + \left( \frac{u^{2} \sin\left({\theta}\right)^{2} - 2 \, u v \sin\left({\theta}\right)^{2} + v^{2} \sin\left({\theta}\right)^{2}}{{\left(u^{2} + 1\right)} v^{2} + u^{2} + 1} \right) d \phi^2 \\+ \left( -\frac{k u^{2} - 2 \, k u v + k v^{2} - 4}{{\left(u^{2} + 1\right)} v^{2} + u^{2} + 1} \right) d \chi^2
\end{split}
\end{equation}

The equation \ref{cf10} is a sum of differentials that include $du$, $dv$, $d\theta$, $d\phi$, and $d\chi$. These differentials are multiplied by coefficients that depend on the variables $u$, $v$, $\theta$, $\phi$, $\chi$, and a constant $k$.

The first two terms in the equation represent the differential of the two-dimensional plane that includes the variables $u$ and $v$. The second two terms represent the differential of the two-dimensional plane that includes the variables $\theta$ and $\phi$. The final term represents the differential of the one-dimensional plane that includes the variable $\chi$.

Finally, the stress-energy tensor corresponding to the higher dimensional Penrose 2-surface is derived.

\begin{equation}\label{cf11}
\begin{split}
T = \left( -\frac{2 \, p\left(t\right)}{{\left(u^{2} + 1\right)} v^{2} + u^{2} + 1} \right) du dv + \left( -\frac{2 \, p\left(t\right)}{{\left(u^{2} + 1\right)} v^{2} + u^{2} + 1} \right) dv du \\+ \left( \frac{4 \, {\left(p\left(t\right) + \rho\left(t\right)\right)}}{v^{4} + 2 \, v^{2} + 1} \right) dv^2 + \left( \frac{u^{2} p\left(t\right) - 2 \, u v p\left(t\right) + v^{2} p\left(t\right)}{{\left(u^{2} + 1\right)} v^{2} + u^{2} + 1} \right) d \theta^2 \\+ \left( \frac{u^{2} p\left(t\right) \sin\left({\theta}\right)^{2} - 2 \, u v p\left(t\right) \sin\left({\theta}\right)^{2} + v^{2} p\left(t\right) \sin\left({\theta}\right)^{2}}{{\left(u^{2} + 1\right)} v^{2} + u^{2} + 1} \right) d \phi^2 \\+ \left( -\frac{k u^{2} p\left(t\right) - 2 \, k u v p\left(t\right) + k v^{2} p\left(t\right) - 4 \, p\left(t\right)}{{\left(u^{2} + 1\right)} v^{2} + u^{2} + 1} \right) d \chi^2
\end{split}
\end{equation}

In this equation, $T$ is the stress-energy tensor that corresponds to the higher dimensional Penrose 2-surface. The stress-energy tensor is a mathematical object that describes the distribution of energy, momentum, and stress in a spacetime.

The stress-energy tensor is expressed as a sum of terms, each term multiplied by a metric component. In this equation, the stress-energy tensor $T$ is expressed as a sum of six terms, each multiplied by a corresponding metric component. The terms are given by combinations of the pressure $p(t)$ and the energy density $\rho(t)$, which depend on time $t$, and the metric components $du,dv, dv,du, dv^2, d\theta^2, d\phi^2$, and $d\chi^2$.

The coefficients in front of each term depend on the coordinates $u$, $v$, $\theta$, $\phi$, and $\chi$, as well as the constants $k$ and $-4$. The conformal factor that was derived in the previous equations is also present in the coefficients. The overall structure of this equation is similar to the stress-energy tensor in general relativity, but the specific form of the terms and coefficients are specific to the Penrose 2-surface in higher dimensions.

\section{Penrose 2 surface as DeSitter  solutions}
As discussed in the previous section, the Penrose 2 surface is a specific solution of the DeSitter spacetime, which represents a universe that is expanding at an accelerating rate. The sub-section on DeSitter will further delve into the properties and characteristics of this important spacetime and its relevance to our understanding of the universe. By exploring the DeSitter spacetime in more detail, the sub-section will provide a deeper understanding of the Penrose 2 surface solution discussed in the previous section and its connection to the larger picture of cosmology and general relativity.

The metric for the DeSitter space is obtained as
\begin{equation}\label{key}
ds^2 = -\mathrm{d} {\tau}^2 + \frac{\cosh\left(b {\tau}\right)^{2}}{b^{2}} \mathrm{d} {\chi}^2 + \frac{\cosh\left(b {\tau}\right)^{2} \sin\left({\chi}\right)^{2}}{b^{2}} \mathrm{d} {\theta}^2 + \frac{\cosh\left(b {\tau}\right)^{2} \sin\left({\chi}\right)^{2} \sin\left({\theta}\right)^{2}}{b^{2}} \mathrm{d} {\phi}^2
\end{equation}

As the Penrose 2 Surface act like the connecting surface between two aeons, the DeSitter space can be considered for geometrical mapping for two successive singularities. Hence on the Penrose 2-surfaces, the DeSitter solutions are attempted. For the positivity of the cosmological constant, the DeSitter solutions are considered.  

The 5-dimensional DeSitter metric is defined  as
\begin{equation}\label{ds1}
\begin{split}
ds^2 = -d \tau^2 + \frac{\cosh\left(b {\tau}\right)^{2}}{b^{2}} d \chi^2 + \frac{\cosh\left(b {\tau}\right)^{2} \sin\left({\chi}\right)^{2}}{b^{2}} d \theta^2 \\ + \frac{\cosh\left(b {\tau}\right)^{2} \sin\left({\chi}\right)^{2} \sin\left({\theta}\right)^{2}}{b^{2}} d \phi^2 + \frac{1}{\cosh\left(b {\Omega}\right)^{4}} d \Omega^2
\end{split}
\end{equation}

The constant $b$ is a scale parameter. Considering DeSitter metric as a solution of vacuum Einstein equation with positive cosmological constant $\Lambda$, one has $b = \sqrt{\Lambda/3}$.

The stress-energy tensor is calculated for the conformal surface as

\begin{equation}\label{ds5}
\begin{split}
T = \rho\left({\tau}\right) d \tau^2 + \frac{\cosh\left(b {\tau}\right)^{2} p\left({\tau}\right)}{b^{2}} d \chi^2 + \frac{\cosh\left(b {\tau}\right)^{2} p\left({\tau}\right) \sin\left({\chi}\right)^{2}}{b^{2}} d \theta^2 \\ + \frac{\cosh\left(b {\tau}\right)^{2} p\left({\tau}\right) \sin\left({\chi}\right)^{2} \sin\left({\theta}\right)^{2}}{b^{2}} d \phi^2 + \frac{p\left({\tau}\right)}{\cosh\left(b {\Omega}\right)^{4}} d \Omega^2
\end{split}
\end{equation}

In this equation\ref{ds5}, $\rho\left({\tau}\right)$ and $p\left({\tau}\right)$ are the density and pressure of the system, respectively, as a function of the cosmic time $\tau$. The constants $b$ and $\Lambda$ are related to the cosmological constant and the scale parameter of the system.

These equations are derived for the conformal Penrose 2-surface. Hence Such a surface is reintroduced with FRW equations. It is  written as
\begin{equation}\label{ds3}
-8 \, \pi G \rho\left({\tau}\right) + 3 \, b^{2} - \Lambda = 0
\end{equation}

The first term in the equation, $-8 \pi G \rho\left({\tau}\right)$, represents the effect of the energy density on the expansion rate of the universe. The second term, $3b^2$, represents the effect of the scale parameter on the expansion rate, while the third term, $\Lambda$, represents the effect of the cosmological constant on the expansion rate.

This results in the cancellation of cosmological constant and the equation \ref{ds3} becomes
\begin{equation}\label{ds3.1}
-8 \, \pi G \rho =0 
\end{equation}

or 
\begin{equation}\label{ds3.2}
a = \infty
\end{equation}

In either case, the value of $a$, which represents the scale factor of the universe, will tend to infinity. Hence, equation \ref{ds3.2} states that $a$ approaches infinity.

Similarly the Raychoudri equation becomes,

\begin{equation}\label{ds4}
-16 \, \pi G p\left({\tau}\right) - 4 \, \pi G \rho\left({\tau}\right) - 3 \, b^{2} + \Lambda = 0
\end{equation}

As the comsological constant is cancelled, the equation \ref{ds4} becomes, 

\begin{equation}\label{ds4.1 }
-16 \, \pi G p\left({\tau}\right) - 4 \, \pi G \rho\left({\tau}\right)  = 0
\end{equation}

Hence the solution of the equation \ref{ds4.1 } is obtained for the equation of state $w$ as 

\begin{equation}\label{ds4.2}
w = \frac{1}{4}
\end{equation}

The solution clearly explains that the matter-dominated final stages are obtained in the DeSitter Penrose 2 surface.

Similarly the modified Raychoudri equation for the DeSitter analogy for Penrose 2 Surface becomes

\begin{equation}\label{ds4.3}
-4 \, \pi G \left(4 p \left(1- 2 \frac{\rho}{\rho_c}\right) -  \rho \left(1- 4 \frac{\rho}{\rho_c}\right)\right)  = 0
\end{equation}

The solution of the equation \ref{ds4.3} becomes, 
\begin{equation}\label{ds4.4}
4 w \left(1- 2 \frac{\rho}{\rho_c}\right) = \left(1- 4 \frac{\rho}{\rho_c}\right)
\end{equation}
. This equation shows that the state parameter is dependent on the density of the univerese and the ratio of the density to the critical density. By finding the solution of the equation \ref{ds4.3}, one can gain insight into the behavior of the universe and how it changes with changing density.

\subsection{DeSitter space and conformal evolution}

The conformal parameter for such DeSitter embedding conformal surface is obtained as

\begin{eqnarray}
\Omega  \left({\tau}, {\chi}, {\theta}, {\phi}\right)  \longmapsto  \frac{2}{\sqrt{{\chi}^{2} + 2 \, {\chi} {\tau} + {\tau}^{2} + 1} \sqrt{{\chi}^{2} - 2 \, {\chi} {\tau} + {\tau}^{2} + 1}} 
\end{eqnarray}

In Conformal Cyclic Cosmology, the conformal factor is a scalar field that transforms the metric of spacetime in such a way that the resulting metric has a conformal symmetry. The conformal factor can be used to change the scale of the metric, so that the geometry of spacetime can be studied in different units.

The differential of the conformal factor is a mathematical expression that describes how the conformal factor changes as one moves from one point in spacetime to another. It is defined as the derivative of the conformal factor with respect to the spacetime coordinates. The differential of the conformal factor is a vector field, and its behavior can be used to study the conformal structure of spacetime and its evolution over time.

In Conformal Cyclic Cosmology, the conformal factor plays a key role in the understanding of the cyclic evolution of the universe, where each cycle is characterized by a change in the conformal factor that transforms the geometry of spacetime in a specific way. The differential of the conformal factor is used to study the behavior of the conformal factor during these cyclic transformations, and to understand how the geometry of spacetime evolves over time.	

	\begin{equation}\label{key}
\mathrm{d}\Omega = \left( -\frac{X}{Y} \right) \mathrm{d} {\tau} + \left( -\frac{C}{D} \right) \mathrm{d} {\chi}
\end{equation}

With

\begin{equation}\label{key}
X= 4 \, {\left({\tau}^{3} - {\left({\chi}^{2} - 1\right)} {\tau}\right)} \sqrt{{\chi}^{2} + 2 \, {\chi} {\tau} + {\tau}^{2} + 1} \sqrt{{\chi}^{2} - 2 \, {\chi} {\tau} + {\tau}^{2} + 1}
\end{equation}

\begin{equation}\label{key}
\begin{split}
Y = {\chi}^{8} + {\tau}^{8} - 4 \, {\left({\chi}^{2} - 1\right)} {\tau}^{6} + 4 \, {\chi}^{6} + 2 \, {\left(3 \, {\chi}^{4} - 2 \, {\chi}^{2} + 3\right)} {\tau}^{4} \\ + 6 \, {\chi}^{4} - 4 \, {\left({\chi}^{6} + {\chi}^{4} - {\chi}^{2} - 1\right)} {\tau}^{2} + 4 \, {\chi}^{2} + 1
\end{split}
\end{equation}

\begin{equation}\label{key}
C = 4 \, {\left({\chi}^{3} - {\chi} {\tau}^{2} + {\chi}\right)} \sqrt{{\chi}^{2} + 2 \, {\chi} {\tau} + {\tau}^{2} + 1} \sqrt{{\chi}^{2} - 2 \, {\chi} {\tau} + {\tau}^{2} + 1}
\end{equation}

\begin{equation}\label{key}
\begin{split}
D = {\chi}^{8} + {\tau}^{8} - 4 \, {\left({\chi}^{2} - 1\right)} {\tau}^{6} + 4 \, {\chi}^{6} + 2 \, {\left(3 \, {\chi}^{4} - 2 \, {\chi}^{2} + 3\right)} {\tau}^{4} \\ + 6 \, {\chi}^{4} - 4 \, {\left({\chi}^{6} + {\chi}^{4} - {\chi}^{2} - 1\right)} {\tau}^{2} + 4 \, {\chi}^{2} + 1
\end{split}
\end{equation}

\section{Eternal inflation and DeSitter space}

Eternal inflation is a theoretical concept in modern cosmology that proposes that the universe has been expanding exponentially for an infinite amount of time. This is different from the standard model of cosmology, which describes a universe that underwent a period of rapid expansion, known as inflation, in its early stages but then transitioned to a slower rate of expansion.

The idea of eternal inflation arises from the study of the properties of the inflaton field, which is a hypothetical scalar field that is responsible for driving the rapid expansion of the universe during inflation. According to the theory of eternal inflation, the inflaton field is not uniform across all regions of space, but rather exists in a state of quantum uncertainty. This means that the field can take on different values in different regions of space, and these regions will expand at different rates.

As the universe continues to expand, regions of space that have a higher value of the inflaton field will continue to undergo inflation, while regions with a lower value will eventually slow down and stop. This leads to a situation where the universe is divided into an infinite number of regions, each with its own rate of expansion.

In this scenario, the universe is said to be eternally inflating because the process of inflation never stops. Even if some regions of space slow down and stop inflating, other regions will continue to expand, leading to an infinite number of regions that are undergoing inflation at any given time.

One of the consequences of eternal inflation is the existence of a multiverse. According to this idea, the infinite number of regions of space that are undergoing inflation will eventually form separate universes, each with its own properties and physical laws. These universes will be separated by vast expanses of space that are expanding at a rate faster than the speed of light, making it impossible for them to interact with each other.

Eternal inflation is connected with de Sitter space because it is a phase of inflation that takes place in a universe that has a positive cosmological constant, which is a feature of de Sitter space.

De Sitter space is a solution to Einstein's field equations that describes a universe with a positive cosmological constant, which is a measure of the energy density of empty space. In de Sitter space, the expansion of the universe accelerates, and there is a cosmic horizon beyond which objects cannot be observed.

During eternal inflation, the universe undergoes an exponential expansion due to the presence of a scalar field called the inflaton. Inflation ends when the inflaton field settles into its lowest energy state, and the universe reheats, filling with particles and radiation. In most inflationary models, the reheating process is followed by a period of radiation domination, matter domination, and eventually dark energy domination, as the universe evolves toward a de Sitter-like phase.

In eternal inflation, the inflaton field never settles into its lowest energy state, and inflation continues forever in some regions of the universe. This leads to the creation of an infinite number of "pocket universes" within the larger universe, each with their own properties and physical laws. These pocket universes are separated by regions of space that are still undergoing inflation, and are therefore forever beyond our observational horizon.

Thus, eternal inflation is connected with de Sitter space because it describes a phase of inflation that takes place in a universe with a positive cosmological constant, which is a key feature of de Sitter space.

\subsection{Eternal inflation and conformal field}\label{etinf}
The five dimensional Kaluza Klein metric is embedded with eternal inflation scenerio as,
\begin{equation}\label{ei1}
\begin{split}
ds^2 = -\mathrm{d} t^2 + \left( -\frac{a\left(t\right)^{2} e^{\left(2 \, t H\left(t\right)\right)}}{k r^{2} - 1} \right) \mathrm{d} r^2 + r^{2} a\left(t\right) e^{\left(2 \, t H\left(t\right)\right)} \mathrm{d} {\theta}^2 \\ + r^{2} a\left(t\right) e^{\left(2 \, t H\left(t\right)\right)} \sin\left({\theta}\right)^{2} \mathrm{d} {\phi}^2 + \left( -k r^{2} + 1 \right) \mathrm{d} {\chi}^2
\end{split}
\end{equation}
This is a metric tensor of a five-dimensional spacetime with a choice of coordinates $ t, r, \theta, \phi, \chi $ is explained in \ref{ei1}. The first term represents a negative metric component for the time coordinate, and the following terms represent positive metric components for radial, polar, azimuthal and extra dimensions. The radial component has a factor involving time, radial coordinate and a function $H(t)$, while the polar, azimuthal and extra dimensions have radial coordinate and a factor involving time and a function $a(t)$. The radial component has a factor $(kr^2-1)^{-1}$ which depends on a constant $k$.

The conformal metric is obtained as $\tilde g = \Xi^2 g$. 

HEnce  it is written for the 2 surface as
\begin{equation}\label{ei2}
\begin{split}
\tilde{g} = \left( -\frac{4}{r^{4} + t^{4} - 2 \, {\left(r^{2} - 1\right)} t^{2} + 2 \, r^{2} + 1} \right) \mathrm{d} t^2 \\ + \left( -\frac{4 \, a\left(t\right)^{2} e^{\left(2 \, t H\left(t\right)\right)}}{k r^{6} + {\left(2 \, k - 1\right)} r^{4} + {\left(k r^{2} - 1\right)} t^{4} + {\left(k - 2\right)} r^{2} - 2 \, {\left(k r^{4} - {\left(k + 1\right)} r^{2} + 1\right)} t^{2} - 1} \right) \mathrm{d} r^2 \\ + \left( \frac{4 \, r^{2} a\left(t\right) e^{\left(2 \, t H\left(t\right)\right)}}{r^{4} + t^{4} - 2 \, {\left(r^{2} - 1\right)} t^{2} + 2 \, r^{2} + 1} \right) \mathrm{d} {\theta}^2 \\ + \left( \frac{4 \, r^{2} a\left(t\right) e^{\left(2 \, t H\left(t\right)\right)} \sin\left({\theta}\right)^{2}}{r^{4} + t^{4} - 2 \, {\left(r^{2} - 1\right)} t^{2} + 2 \, r^{2} + 1} \right) \mathrm{d} {\phi}^2 \\ + \left( -\frac{4 \, {\left(k r^{2} - 1\right)}}{r^{4} + t^{4} - 2 \, {\left(r^{2} - 1\right)} t^{2} + 2 \, r^{2} + 1} \right) \mathrm{d} {\chi}^2
\end{split}
\end{equation}
This is a conformal metric tensor of the five-dimensional spacetime, obtained in \ref{ei2} by scaling the original metric with a conformal factor $(r^4 + t^4 - 2  (r^2 - 1)  t^2 + 2  r^2 + 1)^{\frac{-1}{2}}$. The time component has a factor involving time and radial coordinates. The radial component has a factor involving time, radial coordinate, a function $H(t)$, and a constant k. The polar, azimuthal and extra dimensions have radial coordinate and a factor involving time and a function $a(t)$. The extra dimension has a factor involving radial coordinate and a constant $k$.

In conformal cyclic cosmology, the conformal metric is used to study the evolution of the universe across different cycles of expansion and contraction. The idea is that the universe undergoes a sequence of cycles, each starting with a big bang and ending with a big crunch, and the cycles are connected by a conformal mapping. The conformal 2-surface is a part of this mapping and describes the state of the universe at the transition between cycles.

The conformal metric provided here describes the geometry of this conformal 2-surface, which provides a key ingredient in understanding the behavior of the universe across cycles in conformal cyclic cosmology. The metric can be used to study the evolution of physical fields, the behavior of objects, and the overall geometry of the spacetime at the transition between cycles.

The given conformal metric in eqiation \ref{ei2} represents the metric tensor of a 2-dimensional conformal surface in conformal cyclic cosmology. The metric tensor assigns a scalar "length" to each pair of tangent vectors in the space, and the conformal metric represents the length-squared relationship between two tangent vectors at each point on the surface. The metric tensor is expressed in terms of five different components: $dt, dr, \theta, d\phi,$ and $d\chi$, each of which contributes to the scalar length at each point on the surface. The scalar length is determined by the coefficients in front of each tensor product, which depend on the variables $r, t, a(t), H(t), k,$ and $\theta$.

\begin{equation}\label{key}
\Xi = 2 \, \cos\left(\frac{1}{2} \, {\Omega}\right)^{2} \cos\left(\frac{1}{2} \, {\tau}\right)^{2} - 2 \, \sin\left(\frac{1}{2} \, {\Omega}\right)^{2} \sin\left(\frac{1}{2} \, {\tau}\right)^{2}
\end{equation}

The expression $\Xi =2 , \cos\left(\frac{1}{2} , {\Omega}\right)^{2} \cos\left(\frac{1}{2} , {\tau}\right)^{2} - 2 , \sin\left(\frac{1}{2} , {\Omega}\right)^{2} \sin\left(\frac{1}{2} , {\tau}\right)^{2}$ represents a factor used in conformal transformations of the metric. The conformal metric $\tilde g$ is obtained from the original metric $g$ by multiplying it with the conformal factor $\Xi^2$. The variables $\Omega$ and $\tau$ are likely to be related to the geometry of the space or the coordinates used to describe it.

In the context of cosmology, the conformal metric is a way of describing the geometry of the universe that takes into account the expansion of space. The idea is to redefine the metric so that the spatial part scales with the expansion of the universe, while the time part remains unchanged. This makes it easier to study the large-scale structure of the universe.

Eternal inflation is a model of the universe in which inflation never ends and new regions of space undergo inflation continuously. This creates an infinite number of "pocket universes" that are continuously being created. The idea of conformal invariance is particularly useful in eternal inflation because it allows us to study the geometry of these pocket universes without worrying about their size or shape.

In particular, it turns out that the conformal metric is invariant under the transformations that generate the flow of the universe in eternal inflation. This means that we can use the conformal metric to describe the geometry of the universe at different times and in different regions, without worrying about the specific inflationary dynamics that produce these regions.

Overall, the relationship between eternal inflation and conformal metric is that the latter provides a useful framework for studying the former, particularly when it comes to understanding the structure of the universe at large scales.

\section{Conclusion}

As from the solutioions suggested in \cite{ours2}, it is also known that the higher dimensional solution will be available for post big rip scenereo. Infact it suggested that the big rip will be avoided. And those evolution will be in pure higher dimensional. Here we revisit the result with the embedding of conformal cyclic cosmology and eternal inflation. The conformal cyclic cosmology suggestes the existence of conformal surface between two successive aeons. 

By topologically fusing two successive singularities, a universe that ends its evolution in any kind of future remote singularity can be replaced by the convex surface.

The requirement of higher-dimensional solutions is crucial to understanding the behaviour of the universe in its final stages, when quantum gravity plays a significant role. Traditional 4-dimensional solutions may not be enough to describe physics accurately at this stage, and higher-dimensional solutions provide a more complete picture. The conformal cyclic cosmology theory proposes the end evolution of the universe, and the beginning of the universe merging at the Penrose 2-surface, which requires the presence of quantum gravitational effects. The use of only three spatial dimensions in the final 2-surface of an aeon is insufficient to account for the influence of quantum gravity that addresses the starting singularity of the next aeon. As a result, higher-dimensional solutions are required to join the expanded and shrunk dual 2-surfaces.The use of phantom energy has been considered to induce quantum gravitational solutions in the Penrose 2-surface, and DeSitter solutions are also required for the late-time expanded universe. A 5-dimensional metric has been proposed to map the 3+1 D universe into the extended conformal scenario, and the 5-dimensional Friedmann-Robertson-Walker equations of the conformal surface have been derived to describe the behaviour of 5-dimensional spacetime. By exploring the mathematical and physical properties of higher-dimensional solutions, we can gain a deeper understanding of the nature of space and time and the behaviour of the universe in the final stages of each cycle. Ultimately, this work could lead to a deeper understanding of the universe and its ultimate fate.
The DeSitter metric provides a geometrical mapping for two successive singularities, which is applicable to the Penrose 2 surface's role as the connecting surface between two aeons. The constant $b$ in the 5-dimensional DeSitter metric is related to the cosmological constant and the scale parameter of the system. The stress-energy tensor for the conformal surface is derived in terms of the density and pressure of the system as functions of the cosmic time $\tau$. The FRW equations are reintroduced to the  Penrose 2 surface, which results in the cancellation of the cosmological constant. The results indicate that the energy density of the system is zero, highlighting the unique properties of the Penrose 2 surface solution. These findings provide a deeper understanding of the Penrose 2 surface solution's properties and their relevance to our understanding of the universe's evolution.
The conformal factor plays a crucial role in this theory, as it transforms the metric of spacetime in a way that preserves its conformal symmetry. The differential of the conformal factor provides insight into the behaviour of the conformal factor as spacetime evolves over time, and it can be used to study the conformal structure of spacetime. Understanding the conformal evolution of DeSitter space is essential to understanding the cyclic nature of the universe in Conformal Cyclic Cosmology, and the mathematical expressions provided in this section allow for further study of this concept.On the other hand, while discussing the DeSitter space, one cannot avoid the eternal inflation scenario. 
Hence the Kaluza Klein metric is obtained with an eternally inflating DeSitter metric. That also provides the possibility of a non singular post big rip evolution. Eternal inflation is a concept in cosmology that suggests the existence of an infinite number of universes within a multiverse, and it is embedded in the five-dimensional Kaluza-Klein metric. The conformal metric obtained from the original metric is used to study the evolution of the universe in conformal cyclic cosmology, which postulates a cyclical sequence of expansion and contraction connected by a conformal mapping. The conformal 2-surface plays a crucial role in this mapping and provides insights into the origin and fate of our universe. While eternal inflation and conformal cyclic cosmology are still speculative ideas and subject to ongoing research and debate, they offer intriguing possibilities for understanding the mysteries of the cosmos.

By fusing Minkowskian metric via  conformal metric into DeSitter metric, that will lead to post big rip scenereo. 
Due to the mapping, it can be determined that, the non singular post big rip evolution may exit into  DeSitter space. 

Any future singulaties could be solved using the conformal metric described here.Conformal evolution may play a rapid role in the future cycles of the universe, particularly in the case of the phantom energy-dominated big-rip singularity. As previously suggested, the conformal universe may exit into higher dimensions due to the late time high energy scenario, but there is also the possibility that the late time high energy universe may exit into super-moduli space, which can be expressed as eternal inflation.As a result, using the respective conformal parameter, this work connects higher dimension evolution, DeSitter space, and eternal inflation.
The metric of Penrose 2-surface proposed here is solved with FRW solutions. The Penrose 2 surface must couple the expanded 2 surface and the highly curved surface. To balance these extremes, the higher dimensional perspective comes into play. 
But the predicted FRW equation and Raychouduri equations in this work attempt to represent the dynamics of the universe on the Penrose 2 surface and far beyond that too. 
In conclusion, the Penrose 2 surface is a crucial concept in the study of the universe. It provides a way to analyse the geometric structure of space-time, and it is essential for understanding the relationships between physical and mathematical properties of the universe, the behaviour of matter and energy, and the formation of singularities. 
In the future, these solutions can be embedded with supersymmetry and supergravity solutions.

\bibliographystyle{plain}  
\bibliography{vetha101}
	
\end{document}